\documentclass[11pt,twoside]{article}

\usepackage{asp2014}

\aspSuppressVolSlug
\resetcounters

\begin{document}

\title{Crossmatching variable objects with the \textit{Gaia} data}

\author{Lorenzo~Rimoldini,$^1$ Krzysztof~Nienartowicz,$^2$ Maria~S\"{u}veges,$^1$ Jonathan~Charnas,$^1$ Leanne~P.~Guy,$^1$ Gr\'{e}gory~Jevardat~de~Fombelle,$^2$ Berry~Holl,$^1$ Isabelle~Lecoeur-Ta\"{i}bi,$^1$ Nami~Mowlavi,$^3$ Diego~Ord\'{o}\~{n}ez-Blanco,$^1$ and Laurent~Eyer$^3$
\affil{$^1$Department of Astronomy, University of Geneva, Chemin d'Ecogia 16, CH-1290 Versoix, Switzerland; \email{Lorenzo.Rimoldini@unige.ch}}
\affil{$^2$SixSq, Rue du Bois-du-Lan 8, CH-1217 Geneva, Switzerland}
\affil{$^3$Department of Astronomy, University of Geneva, Chemin des Maillettes 51, CH-1290 Versoix, Switzerland}}

\paperauthor{Lorenzo~Rimoldini}{Lorenzo.Rimoldini@unige.ch}{}{University of Geneva}{Department of Astronomy}{Versoix}{Canton of Geneva}{CH-1290}{Switzerland}
\paperauthor{Krzysztof~Nienartowicz}{Krzysztof.Nienartowicz@unige.ch}{}{SixSq}{}{Geneva}{Canton of Geneva}{CH-1217}{Switzerland}
\paperauthor{Maria~S\"{u}veges}{sueveges@mpia.de}{}{Max Planck Institute for Astronomy}{Galaxies and Cosmology Department}{Heidelberg}{}{D-69117}{Germany}
\paperauthor{Jonathan~Charnas}{Jonathan.Charnas@unige.ch}{}{University of Geneva}{Department of Astronomy}{Versoix}{Canton of Geneva}{CH-1290}{Switzerland}
\paperauthor{Leanne~P.~Guy}{Leanne.Guy@unige.ch}{}{University of Geneva}{Department of Astronomy}{Versoix}{Canton of Geneva}{CH-1290}{Switzerland}
\paperauthor{Gr\'{e}gory~Jevardat~de~Fombelle}{Gregory.Jevardat@unige.ch}{}{SixSq}{}{Geneva}{Canton of Geneva}{CH-1217}{Switzerland}
\paperauthor{Berry~Holl}{Berry.Holl@unige.ch}{}{University of Geneva}{Department of Astronomy}{Versoix}{Canton of Geneva}{CH-1290}{Switzerland}
\paperauthor{Isabelle~Lecoeur-Ta\"{i}bi}{Isabelle.Lecoeur@unige.ch}{}{University of Geneva}{Department of Astronomy}{Versoix}{Canton of Geneva}{CH-1290}{Switzerland}
\paperauthor{Nami~Mowlavi}{Nami.Mowlavi@unige.ch}{}{University of Geneva}{Department of Astronomy}{Versoix}{Canton of Geneva}{CH-1290}{Switzerland}
\paperauthor{Diego~Ord\'{o}\~{n}ez-Blanco}{dordonezb@gmail.com}{}{University of Geneva}{Department of Astronomy}{Versoix}{Canton of Geneva}{CH-1290}{Switzerland}
\paperauthor{Laurent~Eyer}{Laurent.Eyer@unige.ch}{}{University of Geneva}{Department of Astronomy}{Versoix}{Canton of Geneva}{CH-1290}{Switzerland}

\begin{abstract}
Tens of millions of new variable objects are expected to be identified in over a billion time series from the \textit{Gaia} mission. Crossmatching known variable sources with those from \textit{Gaia} is crucial to incorporate current knowledge, understand how these objects appear in the \textit{Gaia} data, train supervised classifiers to recognise known classes, and validate the results of the Variability Processing and Analysis Coordination Unit (CU7) within the \textit{Gaia} Data Analysis and Processing Consortium (DPAC). The method employed by CU7 to crossmatch variables for the first \textit{Gaia} data release includes a binary classifier to take into account positional uncertainties, proper motion, targeted variability signals, and artefacts present in the early calibration of the \textit{Gaia} data. Crossmatching with a classifier makes it possible to automate all those decisions which are typically made during visual inspection. The classifier can be trained with objects characterized by a variety of attributes to ensure similarity in multiple dimensions (astrometry, photometry, time-series features), with no need for a-priori transformations to compare different photometric bands, or of predictive models of the motion of objects to compare positions. Other advantages as well as some disadvantages of the method are discussed. Implementation steps from the training to the assessment of the crossmatch classifier and selection of  results are described. 
\end{abstract}

\paragraph{Introduction}

The crossmatch of celestial objects makes it possible to combine complementary information from data collected at various epochs, with different observational and instrumental features (such as wavebands, time sampling, duration, sky coverage, photometric and astrometric accuracy), and also to extract new information by leveraging the synergy among data sets.
At the same time, some of the differences in instrumentation and data taking, convolved with the properties of the objects to crossmatch (herein named targets), can lead to misses and false detections \citep{2007cs........1172G}, which can become numerous as the number of targets grows. Common causes of crossmatch errors include large positional uncertainties, proper motion, variability, blended objects, spurious sources, detector edges or gaps, contamination, noise, etc. 
Variable objects can be more challenging to crossmatch than constant sources, but they also provide additional features which can be exploited to aid in the identification of correct matches.
For each object, we consider multiple characteristics derived from astrometry, photometry, and light curves, in combination with additional information from literature. 
Machine-learning classifiers are convenient tools to handle multi-dimensional tasks, automate the variety of decisions common in visual inspections, and minimize the occurrence of false positives and negatives. 
Supervised classifiers have previously been used for crossmatching catalogues with large positional uncertainties \citep{2012ApJS..203...32R}.
Inspired by this work, we extended the classifier method to make full use of the time series information and applied it to crossmatch variable sources in the \textit{Gaia} data with a selection of surveys, for use in validation and training of variability types.

\paragraph{Method} 

The main steps to crossmatch with a classifier are outlined below and followed by a brief summary of the pros and cons of the method.

\textbf{Neighbours.} 
The first step to crossmatch a set of targets in a data set is to find the corresponding neighbours in another data set within some angular radius from the targets, after making sure that the coordinates of the two data sets are compared in the same reference system, defined with the same equinox, and possibly taking into account the epoch of observation if the displacement by proper motion over time is not negligible (which might imply a search radius much greater than commonly used). 
We searched for neighbours with efficient PostgreSQL queries making use of the Quad Tree Cube sky indexing scheme  \citep{2006ASPC..351..735K}. Our search radius was limited to 5~arcsec, accounting mostly for positional uncertainties of ground-based surveys, as most targets were located in the Large Magellanic Cloud (and thus with negligible proper motion effects).

\textbf{Match criteria.} 
Classification attributes are computed to distinguish matches from non-matches with several criteria from astrometry (angular separation), photometry (e.g., mean brightness, colour), time-series parameters (e.g., central moments and other statistics characterizing the variability), which also incorporate results from literature like periodicity (light curves folded by their most significant periods can be compared effectively, e.g. by the phases of brightness extrema or by a reduced point-to-point scatter).
Depending on the criterion, it can be useful to include values as computed in each data set as well as from their comparison (differences or ratios).
While the classifier should identify correct matches without relying on positions, if proper motion is relevant, its value could be correlated by the classifier with the angular separation from the target and thus reduce the risk of contamination with similar-looking neighbours.

\textbf{Training objects.}
The selection of objects for the training set is one of the most critical phases of supervised classification. To ensure reliable results, a special effort is made to: 
\textit{(i)}~provide a good representation of all match and non-match criteria as a function of variability type and data quality
(if the classifier gives different weights to classes depending on their relative representation, we suggest to use a similar number of training matches and non-matches);
\textit{(ii)}~embed all possible reasons which drive visual inspection-based decisions, including as many challenging cases as possible;
\textit{(iii)}~verify that the misclassification level is low and that the objects among false positives and negatives correspond to acceptable mistakes, or improve the definition of misclassified objects in the training set and iterate until the above-mentioned conditions are met.  
Occasionally, additional dedicated classifiers might be needed to deal with especially difficult cases (e.g., to recover matches from objects initially classified as non-matches).  

\textbf{Optimisation.}
For robust results and to avoid model overfitting, the classifier is optimised by its internal parameters (depending on the method) and by selecting an appropriate subset of the most useful classification attributes \citep[e.g., by forward selection or backward elimination, see][]{Guyon.Elisseeff.Variable.Selection}.
Misclassified training-set objects from the optimised model are then assessed as in item \textit{(iii)} of the training-set selection.

\textbf{Classification.}
Finally, the classifier model is applied to the objects to crossmatch. In the current version, we assume that only one match is associated with each target and vice-versa.
When the \textit{Gaia} observations split sources which are unresolved or blended in other surveys, there is still some chance that the variable object is correctly identified by its variability pattern (unless the system includes multiple variable sources).
Crossmatch results might include multiple match candidates per target and matches associated with multiple targets. 
We decided to select first the highest probability match for each target.
If among the selected matches more than one target is associated with the same match, different options are possible: retain the safest matches (keeping the one with the highest probability and then iterating on the remaining targets for the next highest match probability until there are only single targets per match) or aim at crossmatch completeness (including lower match probabilities but for more targets).
For the \textit{Gaia} data, we chose to base our selection on the reliability of the crossmatch (based on the highest probability), as presented in Fig.~\ref{figO8-4:xm_multimatchtarget}.

\articlefigure[width=0.65\textwidth]{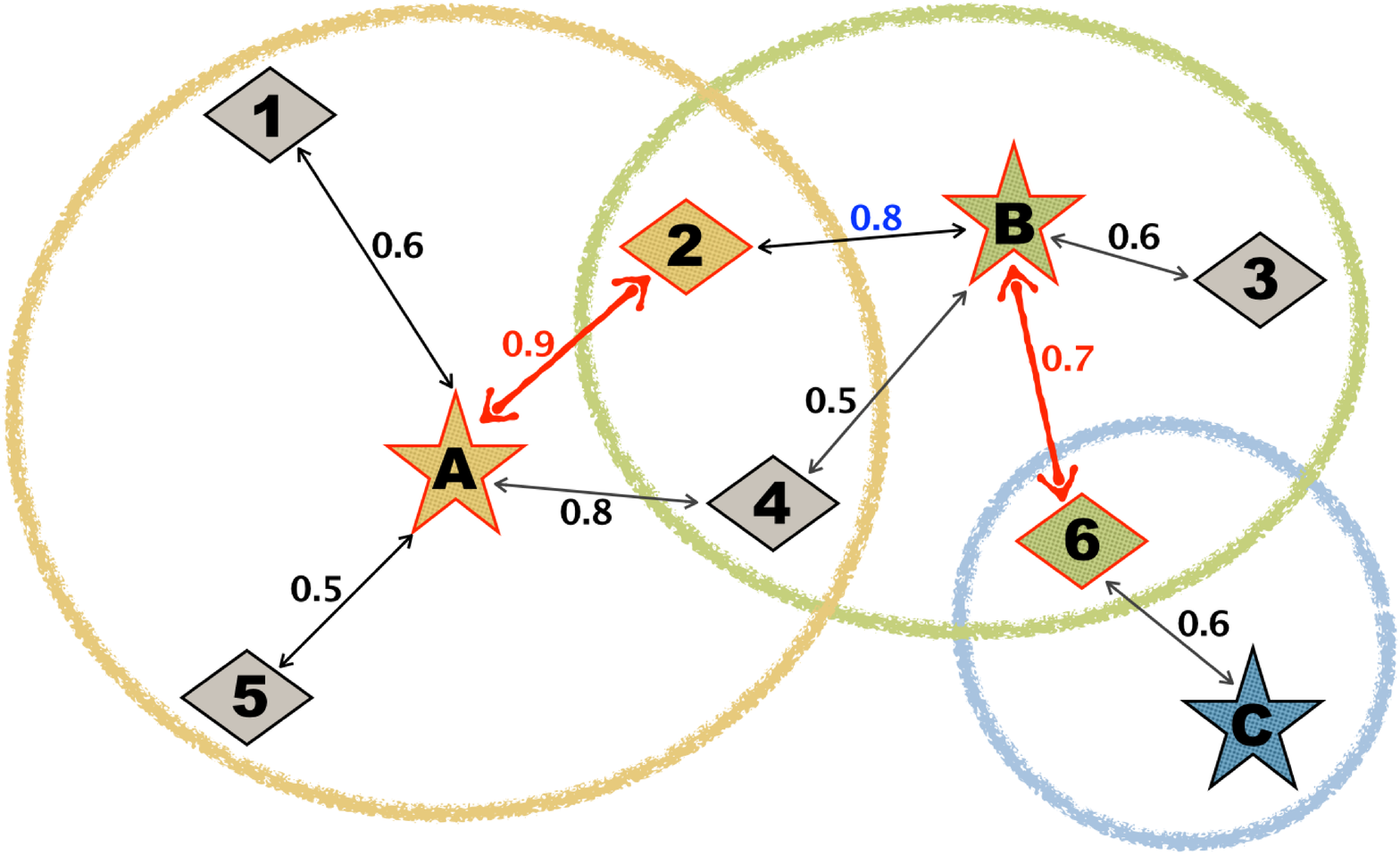}{figO8-4:xm_multimatchtarget}{Probability-based selection of the best-match objects. Targets to crossmatch are denoted by letters (A, B, C) and their match candidates by integer numbers (from 1 to 6), while the respective match probabilities are indicated by the values next to the arrows connecting each object to one or more targets. 
In the cases of multiple matches per target, we select the match-target pairs with the highest probabilites.  
When the same match is associated with two targets, it is assigned to the target with the highest probability, leaving the other target to another match (with the second best probability), if any. In the case depicted here, target~A is associated with object~2, target~B with object~6, and target~C remains unmatched.}

\textbf{Assessment.}
Classification results are assessed by inspecting low-probability non-matches and matches, the farthest matches, the nearest non-matches, and other potential border-line cases. While misclassifications are almost inevitable, the cases which cannot be missed are included in the training set (possibly with additional similar objects) and the steps from the classifier optimisation are iterated until misclassifications are acceptable.  Further diagnostics of the global results, such as the distribution of matches in magnitude- and colour-difference space, can help highlight issues and direct corrective actions (like the selection of new training set objects and/or attributes).

\textbf{Pros and cons.} In summary, the method of crossmatching with a classifier has several advantages with respect to traditional position-based techniques:
\textit{(i)}~the ability to characterize objects by a variety of features to better differentiate the match vs.\ non-match classes and automatically minimize the error rate; 
\textit{(ii)}~robustness of results as the classifier adapts to the data and discovers intrinsic relations: imperfect calibrations do not prevent optimal results, biases caused by artefacts are accounted for (as trained), measurements in different photometric bands can be compared directly without a-priori transformations (as long as the quantities which define them, such as brightness and colour, are included as attributes);
\textit{(iii)}~better performance than a single multi-di\-men\-sional metric, as it does not depend on the accuracy of the components or theoretical expectations;
\textit{(iv)}~selectivity based on variability: if the variability signals of a target and a match candidate are different, the classifier can be taught to consider the pair as a match (e.g., if the signal is only partially sampled) or non-match (e.g., if there is no interest in an eclipsing binary with no measurement in the eclipses);
\textit{(v)}~independence from astrometric details: matches with low positional accuracy or significant proper motion can be identified without knowledge of positional uncertainties or predictive models of their positions (as long as they are within the neighbour search radius);
\textit{(vi)}~the classifier returns a reliability score in the form of an estimate of the probability of matches, which can also be used to set different thresholds depending on the purpose (e.g., a higher threshold for training variability types and a lower one for completeness analyses). 
On the other hand, the main disadvantage of the supervised classifier method is that it depends on the training set (by definition) and it takes time to select training-set objects properly.
As every survey is unique, new classifiers must be trained to crossmatch with different data sets.  Considering the time to visually inspect hundreds of sources for a good training set, the visual confirmation of the best match among the neighbours can be more efficient when the number of crossmatch targets is less than about a thousand.

\paragraph{Results}
The method described herein was applied by means of Random Forest classifiers \citep{Breiman.Random.Forest} to crossmatch known variable objects with the \textit{Gaia} data. Full details of the crossmatch results, crossmatched catalogues, number of matches per catalogue and their sky coverage are presented in \citet{2017arXiv170203295E}. Crossmatch targets covered primarily the region near the LMC, mostly from the OGLE-IV \citep{2012AcA....62..219S,2015AcA....65..233S,2015AcA....65..297S} and the EROS-II \citep{2014A&A...566A..43K,2007A&A...469..387T} surveys. 

\vspace{-0.3cm}
\bibliography{O8-4}  

\begin{thebibliography}{}
\expandafter\ifx\csname natexlab\endcsname\relax\def\natexlab#1{#1}\fi
\expandafter\ifx\csname url\endcsname\relax
  \def\url#1{\texttt{#1}}\fi
\expandafter\ifx\csname urlprefix\endcsname\relax\def\urlprefix{URL }\fi
\providecommand{\eprint}[2][]{\url{#2}}

\bibitem[{{Breiman}(2001)}]{Breiman.Random.Forest}
{Breiman}, L. 2001, Machine Learning, 45, 5

\bibitem[{{Eyer} et~al.(2017)}]{2017arXiv170203295E}
{Eyer}, L., et~al. 2017, (submitted). \eprint{1702.03295}

\bibitem[{{Gray} et~al.(2007){Gray}, {Szalay}, {Budavari}, {Lupton},
  {Nieto-Santisteban}, \& {Thakar}}]{2007cs........1172G}
{Gray}, J., {Szalay}, A., {Budavari}, T., {Lupton}, R., {Nieto-Santisteban},
  M., \& {Thakar}, A. 2007, eprint. \eprint{cs/0701172}

\bibitem[{{Guyon} \& {Elisseeff}(2003)}]{Guyon.Elisseeff.Variable.Selection}
{Guyon}, I., \& {Elisseeff}, A. 2003, J. Machine Learning Res., 3, 1157

\bibitem[{{Kim} et~al.(2014){Kim}, {Protopapas}, {Bailer-Jones}, {Byun},
  {Chang}, {Marquette}, \& {Shin}}]{2014A&A...566A..43K}
{Kim}, D.-W., {Protopapas}, P., {Bailer-Jones}, C.~A.~L., {Byun}, Y.-I.,
  {Chang}, S.-W., {Marquette}, J.-B., \& {Shin}, M.-S. 2014, \aap, 566, A43.
  \eprint{1403.6131}

\bibitem[{{Koposov} \& {Bartunov}(2006)}]{2006ASPC..351..735K}
{Koposov}, S., \& {Bartunov}, O. 2006, in ADASS XV, edited by C.~{Gabriel},
  C.~{Arviset}, D.~{Ponz}, \& S.~{Enrique}, vol. 351 of Astronomical Society of
  the Pacific Conference Series, 735

\bibitem[{{Richards} et~al.(2012){Richards}, {Starr}, {Miller}, {Bloom},
  {Butler}, {Brink}, \& {Crellin-Quick}}]{2012ApJS..203...32R}
{Richards}, J.~W., {Starr}, D.~L., {Miller}, A.~A., {Bloom}, J.~S., {Butler},
  N.~R., {Brink}, H., \& {Crellin-Quick}, A. 2012, Astrophys. J. Suppl. Series,
  203, 32. \eprint{1204.4180}

\bibitem[{{Soszy{\'n}ski} et~al.(2012)}]{2012AcA....62..219S}
{Soszy{\'n}ski}, I., et~al. 2012, Acta Astron., 62, 219. \eprint{1210.1219}

\bibitem[{{Soszy{\'n}ski} et~al.(2015{\natexlab{a}})}]{2015AcA....65..233S}
--- 2015{\natexlab{a}}, Acta Astron., 65, 233. \eprint{1508.00907}

\bibitem[{{Soszy{\'n}ski} et~al.(2015{\natexlab{b}})}]{2015AcA....65..297S}
--- 2015{\natexlab{b}}, Acta Astron., 65, 297. \eprint{1601.01318}

\bibitem[{{Tisserand} et~al.(2007)}]{2007A&A...469..387T}
{Tisserand}, P., et~al. 2007, \aap, 469, 387. \eprint{astro-ph/0607207}

\end{thebibliography}

\end{document}